\documentstyle[sprocl]{article}

\title {Highlight of Dubna-SPIN97 Workshop}
\author{A. Efremov \footnote{ Supported by RFBR Grant No 96-02-17631}}
\address{JINR, Dubna, 141980 Russia}
\date{}
\begin{document}
\maketitle

Processes with polarized particles have been always among the most 
difficult and complicated problems both for experimentalists and theorists.

First, working with polarized targets, experimentalists have to battle with 
thermal "chaos" which trends to break the polarized order. For this one 
needs the liquid helium temperatures. More difficulties, like depolarizing 
resonances, are met in acceleration of polarized particles and in 
controlling a polarized beam.  Second, spin effects are very perfidious: as 
a rule, they are most strong in kinematical regions where the process 
itself is the least probable.

As for the theory, I hardly recall a case when its first prediction was 
correct! As a rule, it was wrong and forced theorists to think more 
fundamentally to repair the theory. This resulted in a deeper understanding 
of particle interaction mechanics. Nevertheless many puzzles like why 
hyperons are produced so strongly polarized or what is the structure of the 
nucleon spin stay yet unsolved during decades. All these problems were the 
subject of those 15 invited and 45 original talk presented at VII-th 
Workshop on High Energy Spin Physics (Dubna-SPIN97) which held in Dubna 
July 7--12, 1997.  What new have we learned there?

\section{The nucleon spin problem}

It is yet a problem although less sharp than ten years ago.  The most 
important thing is that it seems everything OK with the very fundamental 
Bjorken Sum Rule if the whole experimental material and theoretical 
corrections (up to $\alpha_S^3$!) are taken into account \cite{savin}. Also 
it was argued \cite{ioffe}, based on more accurate calculations of QCD Sum 
Rules, that the high twist corrections are much smaller than thought 
earlier:  
$$ 
\Gamma^{\rm tw-4}_{p-n}=-\frac{0.006\pm0.012}{Q^2} 
$$ 
(i.e. $\le2\%$ at $Q^2\ge 5\, GeV^2)$.

The Ellis-Jaffe Sum Rule is definitely violated and the violation is 
especially large (up to $9\sigma$) if one assumes for asymmetry $A(x=0)=0$ 
and $A(x=1)=1$, what has some grounds \cite{savin}. One should have in mind 
however that high twist contributions to singlet $\Gamma_{p+n}$ are 
theoretically rather indefinite and that there is a noticeable difference 
in $\alpha_S$ from E154 and other experiments combined.

Other problems discussed were the radiative and nuclear corrections.  It 
was shown \cite{akushevich} that the former are rather small and for $^3He$ 
do not exceed 5\% in the measured kinematical region.  A new method for 
extracting neutron structure functions from deuteron data was proposed 
\cite{molochkov} based on the Lorentz covariant Bethe-Salpeter equation. 
Though it was shown that the contribution of the antinucleon degrees of 
freedom are suppressed, there are another distinctions with the commonly 
accepted approach and the method should be tested for some concrete 
experimental data.

An open problem in checking the Sum Rules is the small $x$ extrapolation of 
spin--dependent structure functions. In recent years it has become still 
more clear that the Regge--behavior extrapolation is inadequate, owing to 
double log contributions in this region: $\sum_n(\alpha_S\log^2x)^n$.  This 
problem was discussed in the Manaenkov talk \cite{mana} who argued that 
effective ladder diagram summation (with constituent quarks and 
Goldstone-pion steps) changes the power of $x$ obtained from QCD. The 
change is especially big for the isoscalar contribution to the nonsinglet 
part of $g_1$. The author prediction is $g_{1}^{NS,I=0}\propto x^{-0.5}$.

The natural question arises as to what is the reason of Ellis-Jaffe SR 
violation?  Is it a large strange quark contribution or a large gluon 
contribution $\Delta G$?

Some indication for the polarized $s\bar s$ sea comes from $\Lambda$ 
polarization in the target fragmentation region in WA59 experiment 
\cite{kotzinian}.  That the strange sea in the proton is really strange was 
demonstrated by the OBELIX Collaboration who discovered \cite{obelix} that 
the ratio of $\phi\pi^0/\omega\pi^0$ annihilation of $p\bar p$ is about 20 
times as high as the naive OZI prediction and $\phi$'s are dominantly 
produced from the triplet state. This was explained by assuming a strong 
polarized strange sea in the nucleon \cite{ellis}. However, very recently 
the collaboration has shown \cite{nomokonov} that a very similar channel 
with $\phi\eta$ production comes dominantly not from the triplet but from 
singlet state.

On the other hand there are strong arguments \cite{forte} from the analysis 
of existing experimental data in favor of rather large positive $\Delta G$.  
Similar arguments were presented at this workshop too \cite{tokarev}.  So, 
one has to be patient and wait for results of future dedicated experiments. 
These includes the semi--inclusive $J/\psi$ and $\Lambda$ production in DIS 
at COMPASS and HERA \cite{kotzinian,tkabladze,sapozhnikov} and asymmetry in 
direct $\gamma$ production at RHIC \cite{skoro,panebrat}.  Also we heard of 
original proposals for measuring the strangeness content.  These are some 
double polarization observables in $\phi$ photoproduction from the proton 
\cite{titov} and strange particle production $N+N\to N+K+Y$ ($Y=\Lambda$ or 
$\Sigma$) near the threshold or in a collinear kinematics \cite{rekalo}.

The density matrix positivity constrain for the NLO Evolution 
Equation~\cite{teryaev} and the connection of the hadron spin structure with 
the magnetic moments and axial coupling constants was also discussed 
\cite{gerasim,jenkovski}.

About two years ago a possibility to measure an orbital momentum 
contribution to the proton spin was discovered \cite{dvcs}. It is Deeply 
Virtual Compton Scattering (DVCS) process. The leading order QCD 
predictions for the DVCS was reviewed shortly \cite{mueller}. It was 
pointed out that the conformaly covariant OPE possesses the predictive 
power for non-forward two-photon processes in the light-cone dominated 
region.

Study of the spin--dependent fragmentation function by the OPAL 
Collaboration via measuring of vector mesons spin alignment and 
polarization of $\Lambda$'s from jets in $e^+e^-\to Z^0\to 2\,jets$ process 
was also reviewed \cite{bock}.

\section{Single spin asymmetries}

This is another domain full of puzzles. I have mentioned earlier the
inclusive hyperon transversal polarization which yet is the problem for
20 years. Almost so old is the high $p_T$ left-right inclusive pion
asymmetry. It has to be very small from the naive QCD application:
$$
A\propto \frac{m_q}{p_T}\alpha_S
$$
Experimentally, however, as we have learned from the Nurushev review talk 
\cite{nurushev2}, the asymmetry increases with $p_T$ and $x_F$ up to 40\%.  
A new measurement of the $\pi^0$ raw asymmetry at $70\, GeV/c$ and 
$90^\circ$ in the cms was presented \cite{vasiliev}. It is close to zero in 
the region $1.3<p_T<2.5\, GeV/c$  and getting negative in the $2.5<p_T<3.2$ 
GeV/c region. Also the spin measurements in inclusive pion production have 
been carried out in the polarized target fragmentation region. The 
experiment has given an indication of the $x_R$ scaling.  A wider 
investigation of scalings in the pion asymmetry shows however 
\cite{okorokov} that the existing data do not allow one to make a firm 
conclusion on the $x_R$ or $x_T$ scaling. As for me, it seems quite natural 
since the twist-3 QCD predicts a more complicated form of the $x_F$ and 
$p_T$ dependence
$$
A = \frac{Mp_T}{p_T^2+M^2}\Phi(x_F,y)a(x_F)
$$
where M is the polarized hadron mass, $x_F=-u/s,\ y=-t/(s+u)$ and $a(x_F)$ 
is a short range sub-process dependent function calculated in Perturbative 
QCD.

The first observation of transverse handedness in diffractive production of 
pion triples in the process $\pi^-(40\,GeV)+A \to (\pi^-\pi^+\pi^-)+A$ was 
reported \cite{efremov}. It was found that the handedness is rather large 
($10\pm1\%$) and behaves like the transverse polarization, i.e. increases 
with $p_T$ and $x_F$.

A new progress in twist-3 QCD single spin asymmetry was reported 
\cite{boer} It was shown that the so-called "gluonic poles" are reducible 
to a time-reversal odd spin distribution function. The question remains 
however, as to what forbids this pole to appear in physical region of some 
exotic processes like $e^+e^-p\to e^+e^-X$?  Also the anomalous dimension 
calculations which govern the $Q^2$-dependence of the higher twist 
structure and fragmentation functions of the nucleon were presented in  
\cite{belitsky}.

\section{Diffractive and intermediate range spin processes}

A few talks dealt with spin processes in the diffractive region and in 
first place with spin--flip Pomeron. A contribution of that sort follows 
from some dynamic models which take into account the meson--cloud structure 
of hadrons and do not vanish as $s\to \infty$ and $|t|/s\to 0$. It was 
shown \cite{goloskokov} that this effect could be seen in $pp2pp$ 
experiment at RHIC in measuring $A_{NN}$ in the region of the first 
diffractive minimum and also in $A_{LL}$ in diffractive $q\bar q$ 
production in COMPASS and HERA. The effect is especially large for heavy 
quarks and could form an important background in measuring $\Delta G$. The 
interference of the spin-flip contribution with the Coulomb one is very 
peculiar and could serve as a polarimetry effect for RHIC \cite{selyugin}.

The experimental data on $pp$-total cross-sections including the 
spin-depen\-dent parts were analyzed with the goal to determine the 
contribution of spin interactions at high energies \cite{nurushev3}. Based 
on the Regge model with cuts, the energy dependence of such contributions 
are estimated for two spin-dependent terms: 1) the total spin dependent 
term, $\sigma_1$, 2) the spin projection dependent term, $\sigma_2$. The 
estimates show that their contributions to the unpolarized total cross 
section, $\sigma_0$, decrease with energy from several per cent around 
$2\,GeV/c$ to $10^{-2}\%$ around $200\,GeV/c$. There is a clear indication 
that the spin effects are sensitive to the Pomeron intercept. In order to 
pin down such effects, the spin dependent total cross-sections must be 
measured with precision better than $10~\mu b$ at $200\,GeV/c$.

Very interesting observations were reported from the Dubna 
Synchro\-phasatron polarized deuteron and monochromatic neutron beams. The 
behavior of tensor analyzing power $T_{20}(k)$ and polarization transfer 
$\kappa(k)$ versus the nucleon internal momentum $k$ up to $k = 1\, GeV/c$ 
and the initial energy dependence of neutron-proton total cross sections 
$\Delta\sigma_L(np)$ and $\Delta\sigma_T(np)$ in the energy range $1-3.7\, 
GeV$ disagree with all traditional nuclear models, based on low energy $NN$ 
data, but seems to agree with QCD motivated predictions with almost fully 
overlapping nucleons \cite{strunov,sharov}. The experiments are planned to 
be continued with a higher accuracy.

I have to apologize that I cannot touch many talks in intermediate energy 
spin physics coming from ITEP and PNPI accelerators, since I do not feel 
myself competent enough in this region. The only words I want to tell are 
the words of encouraging to people who continue to obtain interesting 
physical results in spite of all difficulties which came down on science in 
Russia.

\section{ Future experiments}

The problems raised by nucleon spin and single spin asymmetries puzzles 
caused an increasing interest in the Spin Physics and initiated a series of 
new dedicated experiments. The programs of some future experiments were 
discussed at the Workshop. We were told the abilities of the COMPASS 
spectrometer for measurements of the gluon contribution, $\Lambda$ 
polarization and charm production \cite{kotzinian,sapozhnikov}, of the 
possibility of HERA-N, which could provide unique information on higher 
twist contributions via single spin asymmetry measurements.  Once the HERA 
proton beam becomes polarized, measurements of the polarized gluon and 
light sea quark distributions in the region of fantastically small $x$ 
might be possible \cite{korotkov}.

Several talks concerned to physics of future RHIC 
experiments\cite{skoro,goloskokov,selyugin,panebrat}.  Especial attention 
was drawn by the $pp2pp$ experiment where the difference in the total cross 
sections as a function of initial transverse spin states, the analyzing 
power, $A_N$, and the transverse spin correlation parameter $A_{NN}$ will 
be measured. Also, the behavior of the analyzing power $A_N$ at RHIC 
energies in the elastic scattering dip region will be studied \cite{guryn}.

The most important problem at RHIC now seems to be the polarimetry of 
polarized beam. Except of a more or less standard method using the 
Coulomb-Nuclear interference \cite{andreeva} which could provide an 
accuracy $\le 5\%$ (after additional precise measurement at AGS), several 
other methods were proposed: asymmetry in elastic $ep$-scattering 
\cite{krechetov} which allows an accuracy $\le 3\%$, the inclusive high 
$p_T$ $\pi^0$ production \cite{nurushev4} and even a spin dependent part of 
synchrotron radiation in a bent crystal \cite{potylitsyn}. From my 
dilettante point of view, at least two of them should be used for mutual 
control.

Last but not least, a very good news was told by Kondratenko 
\cite{kondratenko}:  The Dubna's Nuclotron could be capable to accelerate 
polarized protons. This rises a new hope that the history of spin physics 
in Dubna, told by Nurushev~\cite{nurushev1}, will obtain a new development.

\medskip
In concluding, I would like to thank the JINR Directorate, the Russian 
Foundation for Basic Research (Grant 97-02-26098) and the International 
Organizing Committee of Spin Symposia for financial support of the 
Workshop.

\end{document}